# Graph Contrastive Learning with Multi-Objective for Personalized Product Retrieval in Taobao Search

Longbin Li, Chao Zhang, Sen Li, Yun Zhong, Qingwen Liu, Xiaoyi Zeng
Alibaba Group, Beijing, China
{lilongbin.llb,adamzhangchao,lisen.lisen}@taobao.com

## ABSTRACT

In e-commerce search, personalized retrieval is a crucial technique for improving user shopping experience. Recent works in this domain have achieved significant improvements by the representation learning paradigm, e.g., embedding-based retrieval (EBR) [17] and collaborative filtering (CF) [18, 35]. EBR methods do not sufficiently exploit the useful collaborative signal and are difficult to learn the representations of long-tail item well [38]. Graph-based CF methods [32] improve personalization by modeling collaborative signal within the user click graph. However, existing Graph-based methods ignore user's multiple behaviours, such as click/purchase and the relevance constraint between user behaviours and items. In this paper, we propose a Graph Contrastive Learning with Multi-Objective (GCL-MO) collaborative filtering model, which solves the problems of weak relevance and incomplete personalization in e-commerce search. Specifically, GCL-MO builds a homogeneous graph of items and then optimizes a multi-objective function of personalization and relevance. Moreover, we propose a modified contrastive loss for multi-objectives graph learning, which avoids the mutual suppression among positive samples and thus improves the generalization and robustness of long-tail item representations. These learned item embeddings are then used for personalized retrieval by constructing an efficient offline-to-online inverted table. GCL-MO outperforms the online collaborative filtering baseline in both offline/online experimental metrics and shows a significant improvement in the online A/B testing of Taobao search.

## 1 INTRODUCTION

With the increasing demand of users to purchase on e-commerce platforms, coupled with more and more new products, search engines are beginning to face greater challenges. There are two main goals of e-commerce search engine, one is to increase the platform's Gross Merchandise Volume (GMV), and the other is to satisfy the user's experience. The search engine is a process of people seeking for items and it requires the returned items that have a strong relevance (i.e., text-relevance) with the users queries.

A common practice of search engine is to construct an inverted index table of terms based on item titles for quickly retrieval. However, these approaches [19, 25, 41] are difficult to consider the personal requests and facing the problem of limitation on the number of product sets. For example, a user searches for "pink slim and high waist dress", and the system returns the product titles "yellow plus size dress", "pink dress suitable for tall people", "suitable loose blue dress", the text-based methods retrieve the second item for this query. The text-based methods can retrieve items relevant to the query, but if the returned results are all text-relevant with the query, it is inconvenient for users to choose their favorite style. Furthermore, with billions of products, the number of items in inverted table for each query term is truncated since the system resource is limited, so text-based methods cannot return the full user's favorite item set that are textually relevant to the query.

With the development of deep learning in recent years, many embedding-based retrieval (EBR) methods [2, 15, 17, 20, 23] have emerged, which map user queries and products into the same space. However, EBR methods do not sufficiently exploit the useful collaborative signal [32] and are difficult to learn the representations of long-tail items well [38]. With users historical behaviors, item-base collaborative filtering (CF) [18, 36] algorithms are applied to the search retrieval system. Specifically, the CF method in e-commerce search is a pipeline of "query-item-item", named q2i2i. The q2i process is defined as: when a user searches in e-commerce platform, his/hers historical behaviour items belonging to the same category with the current query will be selected as trigger items. The i2i process is performed with CF methods to construct an item to item inverted index table. Trigger items will look up their similar items in the inverted index table and return the item set as result. However, these methods [18, 36] have several drawbacks, one is that higher-order item similarity relationships cannot be guaranteed, and the other is that it ignores multiple intentions of users.

There have been many works applied graph learning to model the higher-order similarity relationship between items[11, 24, 30]. Although they model similarities between items using users' behaviors, only a single objective is considered in the learning process, such as the common clicks belonging to one user. We call this a single-objective collaborative filtering model, which has several disadvantages in e-commerce search:

**Weak relevance**. In e-commerce search, the relevance between query and result items needs to be satisfied. If only the co-occurrence between items is considered, the constraint of query will be ignored, degrading search relevance. Although users click on these items at the same time, if these items are not relevant to the query, it is hard to consider they are relevant.

**Incomplete personalization**. In e-commerce platforms, items are not only clicked together, but also purchased together. Co-exposure is also a useful relationship between items, even though its signal is weaker than clicks/purchases. Given a user, there are three associations between items, i.e., exposure, click, and purchase relation, while the existing single-objective models do not model them at the same time.

In this paper, a graph contrastive learning with multi-objective (GCL-MO) method is proposed to learn the similarity between items in e-commerce platforms, and the inverted index of items is constructed according to the similarity for the matching stage. GCL-MO learns to directly model the multiple relationships between trigger items and candidate items, which is more direct than learning the similarity between neighbor nodes in the graph structure. Specifically, through the exposure log of the search engine, we get all triggers for each query of the user, as well as a page of impression items for a search. $n$ items will be exposed for one search request, and establish multi-matching relationships with user trigger items (i.e., historical behaviours), including four relationships: relevant, exposure, click, and purchase. Therefore, trigger items can build edge connections with the items on the search exposure page at the first level (objective-level), used for objective construction of graph learning. At the second level (neighbor-level), we build neighbor relationship graph with items of the same category co-clicked by user in recent days, which makes items representations learn better by aggregating common neighbors, especially for long-tail items. Hence, we construct a two-level graph structure, one for learning multiple objectives relationships between items, and one for transferring neighbor relationships between items.

The main contributions of this work are summarized as follows:

- We propose a novel graph contrastive learning with multi-objective (GCL-MO) method to model four relationships (relevance, exposure, click and purchase) between items in e-commerce search. We design a two-layer graph structure to learn their objective-level and neighbor-level relationship.
- We introduce contrastive learning to avoid the learning of graph structure being overly biased towards the hot items, making the representations of long-tail items more generalizable and robust. We propose a modified contrastive loss avoiding the mutual suppression among positive examples.
- The offline qualitative and quantitative analysis, as well as online A/B tests of Taobao search, demonstrate the effectiveness of GCL-MO.

## 2 RELATED WORK
### 2.1 Multi-Objection Optimization
In the field of deep learning, multi-objective optimization has been developed for many years and obtained huge effect. By learning multi-objective, model can improve efficiency and weigh the importance of each objective [4, 7, 26]. A lot of related studies [3, 16, 34] focus on how to model the relations between the multiple objectives (e.g., click, diversity, purchase likelihood). Long et al. [21] combined relevance scores and purchase prediction scores to perform search rankings. Dai et al. [9] generated a mixed label to incorporate the relevance and freshness of candidate items for ranking optimization. Carmel et al. [6] applied the label aggregation method to aggregate labels of training samples with different objectives into a single label, reducing the problem to single-objective learning.

### 2.2 Graph-Based CF Methods
Graph-based learning algorithms have been proposed as a general network representation method. There has been lots of research in this field focusing on designing new embedding algorithms. Line [1] applies factorization methods to approximately factorize the adjacency matrix and preserve both first order and second proximities. Other methods [5, 28, 29] enhance the model's ability of capturing non-linearity in graph by deep learning technology. Due to the extremely large-scale of the graph, [10, 14, 24] use random walk on graph structure to learn efficient node representations. In [39, 40], embeddings of users and items are learned under the supervision of meta-path and meta-graphs, in heterogeneous graphs. These graph learning methods are all optimized in the graph mechanism at the same level and learn with a single objective.

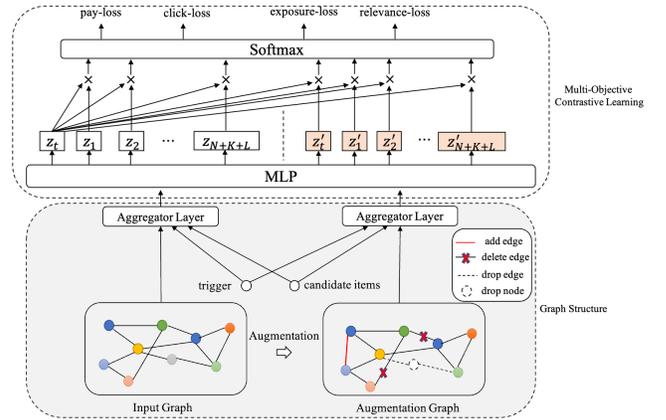

**Figure 1: GCL-MO:Graph Contrastive Learning with Multi-Objective model architecture. Triggers and candidate items are aggregated from original graph and augmented graph obtaining two vectors. In addition to the two vectors $z_t$ and $z'_t$ of trigger, there are $2 \times (N + K + L)$ vectors for candidate items, where $z_i$ represents the vector formed from the aggregation of original graph, $z'_i$ denotes a vector aggregated from augmented graph. $z_t$ computes the inner-product with all other vectors and constructs four contrastive loss functions.**

## 3 PROPOSED METHOD
We first introduce the construction method of our two-level graph structure, then introduce how to perform multi-objective learning on the graph structure, and finally introduce how to use the method of contrastive learning to obtain better representations of items. Figure 1 shows the overview architecture of GCL-MO.

### 3.1 Problem Definition
For user $u$, query $q$, predicted category $c$ of $q$, the collaborative filtering algorithm of e-commerce search is defined as: selecting items belonging to category $c$ in the user's historical behavior sequence $\mathcal{H} = \{i_1, ..., i_T\}$ as triggers, building a function $\mathcal{F}$, which can learn the embeddings for all items and seek the topk similar items for triggers through the nearest neighbor retrieval algorithm.

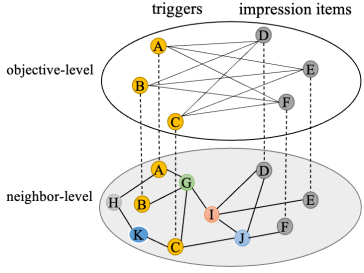

Figure 2: The two-level structure of items relationship.

## 3.2 Graph Structure

We adopt a two-level compositional approach to characterize the connections between items in the graph structure. The structure of the first level represents the objective learning relationship between items, and the structure of the second level represents the neighbor co-occurrence relationship between items, as shown in Figure 2. In this section, we introduce the generation process of the item vector under the two-level graph structure, as shown in Figure 3.

*3.2.1 Objective-Level.* When users search in e-commerce, system will get triggers according to the query, and objective optimization edges are established between triggers and n items exposed under a search process. As shown in Figure 2, at the objective-level, triggers are on the left and all exposed items are on the right.

*3.2.2 Neighbor-Level.* In the field of e-commerce search, relevance is a metric that must be guaranteed, so the same category is used as a constraint when building neighbor relationships. The construction method of neighbor relationship in neighbor-level is that the items under the same category that users have clicked together in recent S days. We take the number of co-occurrences as the edge weight.

*3.2.3 Neighbors Sampling.* We introduce a method of sampling and aggregating neighbor nodes by sampling neighbor nodes only in a local area to obtain a subgraph. For a given node $u_i$, we sample its neighbors according to the probability: $P(v_j|u_i) = \frac{W_{ij}}{\sum_{j \in N(v_j)} W_{ij}}$.

*3.2.4 Aggregator Layer.* GCL-MO mainly learns an aggregation function, which aggregates information from the neighbors of the current node. We try different aggregation functions, finding that aggregation methods have little effect on experimental results. We choose to use the attention method for aggregation.

*3.2.5 Output Layer.* We use the normalized inner product to calculate the similar score. The similar score between item u and item v is defined as: $s_{vu} = <z_v, z_u> / (\|z_v\| \cdot \|z_u\|)$

## 3.3 Multi-Objective Optimization

In this section, we introduce how to perform multi-objective learning in the constructed two-level graph structure. The overall structure of multi-objective graph learning is shown in Figure 4. The trigger will calculate the similarity score with the item under a search exposure page, and then calculate every loss. The function $\phi$ in Figure 4 represents the generation process of the item vector, as shown in Figure 3. To enable GCL-MO to retrieve items that satisfy multiple objectives, we improve its sample construction and propose a new multi-objective loss function.

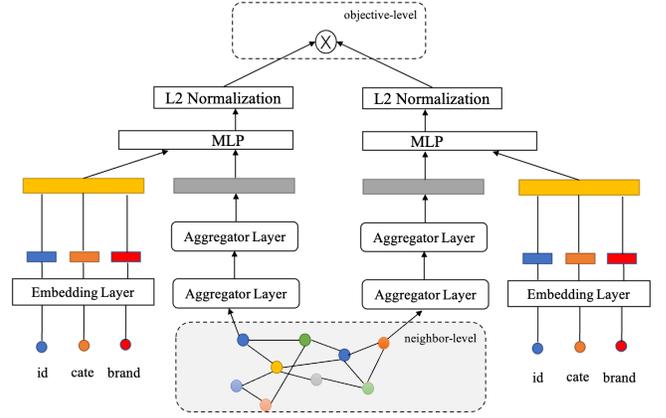

Figure 3: An illustration the generation process of the item vector in GCL-MO model. The source of item's embedding has two parts, one part is the embedding of its ID and other features, and the other is aggregated from the neighbor nodes.

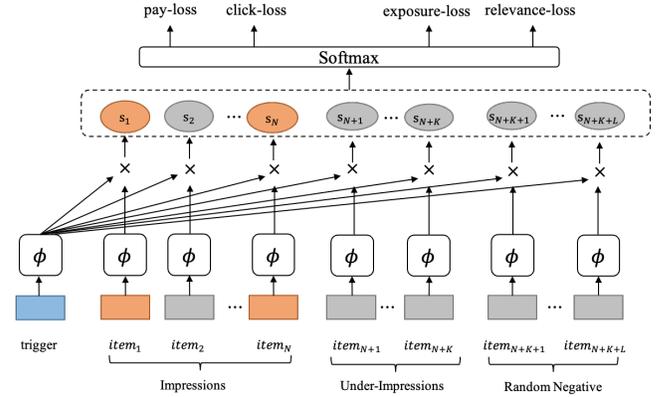

Figure 4: Multi-objective retrieval model GCL-MO with multi-positive training samples.

*3.3.1 Sample Construction.* In a single-objective optimization graph learning model, the similarity relationship between an item and its neighbors is optimized. This model contains only one positive sample, and the optimization goal is to maximize the predicted score of the single positive sample.

Different from the single-positive graph learning model, GCL-MO is trained on page-level training instances with multiple objectives and multiple positive items as shown in Figure 4. A instance of GCL-MO contains all the items exposed in the e-commerce search engine page view, which can directly optimize the relationship between triggers and positive and negative samples. Specifically, in each training instance of GCL-MO, there are several sets of items:

- **Impressions**: N items are exposed in one page, and these exposed items include items clicked or purchased by user.
- **Under-Impressions**: We sampled K items from the unexposed items in the online log system.
- **Random Negatives**: L items that are randomly sampled from the item corpus as negatives.

As shown in Figure 4, for all items in each instance, we construct four binary labels corresponding to the four optimization objectives:

- **Relevance Label**: This label defines that two items are relevant if they are both relevant to the same query. Whether the item is relevant to the query, we use an online relevance estimation model to predict.
- **Exposure Label**: The label is 1 for impressions and 0 for under-impressions and random negatives.
- **Click Label**: The label indicates whether the user clicks, the clicked label is 1, and the unclicked label is 0.
- **Purchase Label**: The label indicates whether the user purchases, the purchase label is 1, and the other is 0.

*3.3.2 Multi-Objective Learning.* GCL-MO has four optimization objectives $O = \{Relevance, Exposure, Click, Purchase\}$. For any objective $o \in O$, given a trigger and candidate items $C = \{item_1, item_2, ..., item_{N+K+L}\}$ which have labels $y^o = \{y_1^o, y_2^o, ..., y_{N+K+L}^o\}$. The number of positive items in $y^o$ is denoted as $|o^+|$. The loss for objective $o$ is defined as follows:

$$y_s^o(i_k|i_t) = \frac{exp(s_{kt}/\tau)}{\sum_{i_j \in C} exp(s_{jt}/\tau)} \quad (1)$$

$$y_k' = min(y_s^o(i_k|i_t) \cdot |o^+|, 1) \quad (2)$$

$$L_o = -\sum_{i_k \in C} y_k^o log(y_k') \quad (3)$$

where $i_k$ is the candidate item $k$ and $i_t$ is trigger item, $s_{kt}$ means the similar score between $i_k$ and $i_t$. We introduce a temperature parameter $\tau$ into softmax to smooth the overall fitted distribution of the training data. The advantage of this operation is to help the model stop optimizing those positive items with excessively large scores. In Equation 2, we multiply the softmax score $y_s^o(i_k|i_t)$ with $|o^+|$ and limit it's value to 1. By this operation, we can minimum value of loss $L_o$ into zero. There are multiple items clicked by the user in the candidate item set $C$, limiting the score $y_k^o$ below 1 can prevent an item with a particularly large score from dominating the entire loss. In other words, in this way, the learning of each positive sample can be balanced. The total loss of GCL-MO is defined by:

$$loss = \sum_{o \in O} w_o L_o \quad (4)$$

where $w_o$ is the loss weight for the objective $o$.

## 3.4 Graph Contrastive Learning

GCL-MO is trained by supervised learning and directly optimizes the trigger and candidate items in a two-level graph structure. Although this method effectively learns the similarity between items, but the neighbor-level nodes are highly dependent on our defined labels in the supervised learning method. It will affect the generalizable and robust representation learning of the model. In e-commerce search, the popularity bias of items is a serious problem, which leads to a large proportion of popular(i.e. head) items in the neighbor nodes of the head item. Contrastive learning has made great progress recently [8, 31], we extend the method of contrastive learning to our multi-objective graph learning to overcome the limitations of the above neighbor-based learning method.

*3.4.1 Graph Data Augmentation.* The purpose of data augmentation is to create novel, plausible data through some transformation without affecting item similarity labels. We introduce two general data augmentations for our neighbor-level graph-structured data, as shown in Figure 1.

**Node dropping.** Given a graph G, node dropping randomly drops some nodes and their connections with probability of dropping. The nodes' dropping probability distribution can be an i.i.d. uniform distribution, or any other distribution.

**Edge perturbation.** This method randomly adds or removes a fraction of the edges. This means that we can disrupt neighbor-level connectivity, and it also means that the semantics of similar items are robust to the variance of edge connectivity patterns.

*3.4.2 Contrastive Loss Function.* We define a contrastive loss function that optimizes the similarity between the trigger and the augmented trigger vector, also and all other candidate item vectors. In the previous section, we have introduced the construction method of four labels. Under the framework of contrastive learning, we need to redefine the four labels. The label of trigger and other items (one augmented trigger representation and $2 \times (N+K+L)$ candidate items) is defined as $y_c^o = \{1, y^o, y^o\}$, where $o \in O$. The first label of $y_c^o$ is setted to 1 for making trigger vector similar to augmented trigger vector. Since we introduce more positive examples in our contrastive learning framework, we propose a new contrastive loss function that can eliminate the mutual suppression of multiple positive examples. We get following inference about multiple-positives classification like Equation 3:

$$L_o = -\sum_{i \in P} log(\frac{exp(s_i)}{\sum_j exp(s_j)}) \quad (5)$$

$$= \sum_{i \in P} log(1 + \frac{\sum_{j \in \{N \cup P_{j \neq i}\}} exp(s_j)}{exp(s_i)}) \quad (6)$$

$$= \sum_{i \in P} log(1 + exp(-s_i + LogSumExp(s_j, j \in \{N \cup P_{j \neq i}\}))) \quad (7)$$

$$\approx \sum_{i \in P} log(1 + exp(-s_i + max(s_j, j \in \{N \cup P_{j \neq i}\}))) \quad (8)$$

where $N$ is the negative examples and $P_{j \neq i}$ is the positive samples other than $i$. We apply LogSumExp function to further change the loss function as Equation 7 show. Since LogSumExp function is approximated as max function, we then approximate LogSumExp function with max function as Equation 8 show. We know from the derivation above that the loss function tries to optimize the difference between each positive sample $i$ and other positive and negative samples. The scores of positive samples are generally larger than those of negative samples, so the result is that the gap between positive samples expands. In order to avoid this result,

we remove other positive samples from LogSumExp function. The contrastive loss of objective o is defined as:

$$y_c^o(i_k|i_t) = \frac{exp(s_{kt}/\tau)}{\sum_{j\in\{N\cup k\}} exp(s_{jt}/\tau)} \quad (9)$$

$$y'_{c,k} = min(y_c^o(i_k|i_t) \cdot |o^+|, 1) \quad (10)$$

$$L_o^c = -\sum_{k\in P} y_{c,k}^o log(y'_{c,k}) \quad (11)$$

Through these two graph augmentation methods, we have made the long-tail items more fully trained. In addition, we removed other positive samples in the denominator of the softmax function to avoid pulling each other between positive samples. In particular, it can prevent long-tail positive samples from being suppressed by other positive samples. Therefore, our proposed contrastive learning method can improve the generalization and robustness of long-tail item representations. In particular, our newly proposed contrastive loss function can be applied to other domains, such as those classification tasks with multiple positive samples.

## 4 EXPERIMENTS

### 4.1 Datasets

We construct our training data from the online click and purchase behavior logs of users collected by Taobao search in 14 days.

For evaluation, we randomly sample 1 million search records of $T+1$ day, including (i) 0.5 million search click records; (ii) 0.5 million search purchase records. The offline i2i(item to item) inverted index engine is consistent with the online environment, which contains about 100 million candidate item.

### 4.2 Evaluation

We evaluate the graph based collaborative filtering performance of models according to these metrics: (1) $recall@K$: whether the target items within the top $K$ items retrieved by triggers; (2) Good rate $P_{good}$: the proportion of items with good relevance in the top $K$ retrieval set. (3) Long-tail rate $P_l$: the proportion of long-tail items in the top $K$ retrieval set. Formally, given the user's click or purchased items as target items $T = \{t_1, t_2, ..., t_N\}$ and the top $K$ retrieval set $I = \{i_1, i_2, ..., i_K\}$ retrieved by triggers, $recall@K$, $P_{good}, P_l$ are defined as follow:

$$Recall@K = \frac{\sum_{i\in I} \mathcal{I}(i \in T)}{N} \quad (12)$$

$$P_{good} = \frac{\sum_{i\in I} \mathcal{R}(i)}{K} \quad (13)$$

$$P_l = \frac{\sum_{i\in I} \mathcal{L}(i)}{K} \quad (14)$$

where $\mathcal{I}(i \in T)$ returns 1 when $i$ is in the target set $T$, otherwise it returns 0, and $\mathcal{R}(i)$ is a relevance indicator that judges whether the item is of good relevance or not to the query. We use our online well-trained relevance model [37] as the relevance indicator to rate the item's relevance. $\mathcal{L}(i)$ is a indicator that judges whether the item is long-tail item(the average daily search exposure in the past 30 days is less than 100).

Table 1: Offline performance of GCL-MO and baseline models, where $K$ is 6,000.

| Model | $Recall@K$ | $Recall_p@K$ | $P_{good}$ |
|---|---|---|---|
| GCL-MO | 0.5732 | 0.5957 | 0.2845 |
| LINE | 0.5472(-2.60%) | 0.5596(-3.61%) | 0.2446(-3.99%) |
| EGES | 0.5537(-1.95%) | 0.5703(-2.54%) | 0.2587(-2.58%) |
| GraphSAGE | 0.5618(-1.14%) | 0.5845(-1.12%) | 0.2687(-1.58%) |
| NGCF | 0.5634(-0.98%) | 0.5854(-1.03%) | 0.2693(-1.52%) |
| NIA-GCN | 0.5647(-0.85%) | 0.5859(-0.98%) | 0.2702(-1.43%) |
| LightGCN | 0.5651(-0.81%) | 0.5864(-0.93%) | 0.2706(-1.39%) |
| DGCF | 0.5655(-0.77%) | 0.5868(-0.89%) | 0.2715(-1.30%) |
| UltraGCN | 0.5661(-0.71%) | 0.5871(-0.86%) | 0.2723(-1.22%) |
| MO-UltraGCN | 0.5695(-0.37%) | 0.5917(-0.40%) | 0.2769(-0.76%) |
| OL GCL-MO | 0.5703(-0.29%) | 0.5923(-0.34%) | 0.2816(-0.29%) |

Table 2: Offline performance of GCL-MO and its ablation models, where $K$ is 6,000.

| Model | $Recall@K$ | $Recall_p@K$ | $P_{good}$ |
|---|---|---|---|
| -relevance loss | 0.5755(+0.23%) | 0.5968(+0.11%) | 0.2413(-4.32%) |
| -exposure loss | 0.5616(-1.16%) | 0.5813(-1.44%) | 0.2834(-0.11%) |
| -click loss | 0.5587(-1.45%) | 0.5848(-1.09%) | 0.2796(-0.49%) |
| -purchase loss | 0.5607(-1.25%) | 0.5841(-1.16%) | 0.2831(-0.14%) |

### 4.3 Experimental Setup

For offline experiments, we compare GCL-MO with the strong baselines LINE [1], EGES [30], GraphSAGE [12], NGCF [32], NIA-GCN [27], LightGCN [13], DGCF [33], UltraGCN [22]. We also train MO-GCL on an one-level structure, named One Level MO-GCL (OL GCL-MO). In addition, in order to verify the effectiveness of multi-objective learning, we also apply multi-objective learning functions on UltraGCN, named MO-UltraGCN.

### 4.4 Offline Experimental Results

*4.4.1 Comparison with Strong Baseline.* As described in Section 4.2, we introduce metrics $Recall@K$ for 1 million search click/purchase records. In the field of e-commerce, user purchase is a very important behavior, so we further report the metrics of $Recall@K$ in the 500,000 search purchase records, which is expressed as $Recall_p@K$. Table 1 reports the comparison results. It can be seen that GCL-MO outperforms the other models on every metric. From the two metrics of $Recall@K$ and $Recall_p@K$, GCL-MO has a larger coverage of user interest than other models. Compared with other strong single-objective methods, such as UltraGCN, $P_{good}$ has a great improvement, which shows that GCL-MO can meet better relevance evaluation. The GCL-MO that directly optimizes the relationship between the trigger and the candidate items performs better than the OL GCL-MO that indirectly optimizes the relationship.

*4.4.2 Effect of Multi-Objective Learning.* To investigate the effectiveness of multi-objective learning, we conducted a series of experiments. The experimental results are shown in Table 2. We can see that the $P_{good}$ of removing either loss function alone decreases.

Especially after removing the relevance loss, $P_{good}$ dropped the most, reaching 4.32%. Even removing the relevance loss can slightly increase $Recall_K$, it does serious damage to $P_{good}$. The performance of GCL-MO also drops significantly when click or purchase losses are removed. Furthermore, we see that exposure loss has a good effect on model performance. The positive samples of the exposure loss are designed as the exposed samples, so the model can learn the pattern of the downstream cascade ranking system. Attributed to Multi-Objectives learning, MO-UltraGCN has a big improvement on $P_{good}$, comparing to UltraGCN as shown in Table 1.

**Table 3: Offline performance of GCL-MO model and other models(without contrastive loss) about $P_l$.**

| Model | $Recall@K$ | $Recall_p@K$ | $P_l$ |
|---|---|---|---|
| GCL-MO | 0.5732 | 0.5957 | 0.1445 |
| GL-MO | 0.5681(-0.51%) | 0.5914(-0.43%) | 0.1364(-0.81%) |
| AP GCL-MO | 0.5703(-0.29%) | 0.5937(-0.20%) | 0.1411(-0.34%) |
| LightGCN | 0.5651(-0.81%) | 0.5864(-0.93%) | 0.1327(-1.18%) |
| DGCF | 0.5655(-0.77%) | 0.5868(-0.89%) | 0.1339(-1.06%) |
| UltraGCN | 0.5661(-0.71%) | 0.5871(-0.86%) | 0.1354(-0.91%) |

*4.4.3 Effect of Contrastive Learning.* We conduct an experiment to investigate the effect of contrastive learning in Graph Learning with Multi-Objective(GL-MO) model. The experimental results are shown in Table 3. AP GCL-MO means All Positives GCL-MO, which does not remove other positive samples in loss function. After introducing the contrastive learning method, all metrics have all been improved to a certain extent. Especially when removing other positive samples in AP GCL-MO loss function, our GCL-MO method further improved $P_l$, which means our proposed contrastive loss can retrieve more long-tail items. We believe that in the field of e-commerce, the neighbors of popular items are almost popular items. If the neighbors are not disturbed, the method based on neighbor aggregation will focus on learning popular items and ignore unpopular items. The method of contrastive learning can enhance the robustness and generalization of the representations of unpopular items without affecting the learning of popular items.

## 4.5 ONLINE A/B TEST

**Table 4: Online performance of GCL-MO and its ablation models.**

| Model | GMV | $P_{good}$ | $P_{click}$ | $P_{purchase}$ |
|---|---|---|---|---|
| GCL-MO | +0.36% | +0.78% | +1.53% | +0.98% |
| GL-MO | +0.14% | +0.37% | +0.65% | +0.37% |
| OL GCL-MO | +0.16% | +0.39% | +0.71% | +0.43% |

It is necessary for us to make a simple description of the environment of our online system. In Taobao search, long-tail items accounted for 91% of the number of items, only have about 13% of exposures and 8% of purchase. Head or hot items accounted for 9%, but have 87% of exposure and 92% of purchase.

Online metrics use (i) Gross Merchandise Volume (GMV) (ii) $P_{good}$: the proportion of items with good relevance in the set of exposed items (iii) $P_{impression}$, $P_{click}$, $P_{purchase}$: the proportions of items retrieved by the GCL-MO model in the set of impression/clicked/purchased items. (iv) $P_{l-impression}$, $P_{l-click}$, $P_{l-purchase}$: $P_{impression}$, $P_{click}$ and $P_{purchase}$ of long-tail items.

As shown in Table 4, our GCL-MO model improves GMV by 0.36% compared to online services, and achieves a $P_{good}$ 0.78% improvement, which shows that GCL-MO can bring greater benefits to the platform, and at the same time can improve the user's browsing experience and enhance the relevance. Since the GCL-MO optimzes the objectives of click and purchase, there are 1.53% and 0.98% improvements in $P_{click}$ and $P_{purchase}$ compared to the online service model. GCL-MO achieve greater GMV and $P_{click}$ compared to GL-MO. All the results show that our proposed GCL-MO model can well improve the relevance and personalization.

Moreover, with the further introduction of contrastive learning methods, GCL-MO can make the learned representations of long-tail items more robust and generalized to a certain extent, and brings great online revenue as shown in Table 5. Compared with the insignificance of the offline experiment results of long-tail items, the online system can help us expand the statistical quantity of long-tail items and make the experimental results more reliable. Especially, $P_{l-impression}$ achieves 3.38% improvement, which means long-tail items get more exposure.

**Table 5: Online performance of GCL-MO in long-tail items.**

| Model | $P_{l-impression}$ | $P_{l-click}$ | $P_{l-purchase}$ |
|---|---|---|---|
| GCL-MO | +3.38% | +2.76% | +1.37% |
| GL-MO | +0.65% | +0.39% | +0.17% |

## 5 CONCLUSION

This paper introduces our Graph Contrastive Learning with Multi-Objective(GCL-MO) model for personalized product retrieval. It mainly solves the problem of weak relevance and incomplete personalization in the collaborative filtering algorithm of e-commerce search engine. GCL-MO adopts the method of multi-objective optimization and constructs four loss functions, including relevance, exposure, click and purchase. Compared with the single-objective graph learning model, GCL-MO has better performance in $P_{good}$, $Recall@K$ metrics. In addition, GCL-MO introduces the method of contrastive learning, which alleviates the generalization representation problem of long-tail items in e-commerce platforms. Offline experiments and online A/B tests verify the effectiveness of GCL-MO. We have deployed the proposed system in Taobao search, serving hundreds of millions of users.